\documentclass[10pt]{article}

\usepackage{amsmath}

\DeclareMathOperator*{\esssup}{ess\,sup} 

\begin{document}

\baselineskip=4.6mm

\makeatletter

\newcommand{\E}{\mathrm{e}\kern0.2pt}
\newcommand{\D}{\mathrm{d}\kern0.2pt}
\newcommand{\RR}{\mathrm{I\kern-0.20emR}}

\def\bottomfraction{0.9}

\title{{\bf No steady water waves of small amplitude are supported by a shear flow
\\ with still free surface}}

\author{Vladimir Kozlov$^a$, Nikolay Kuznetsov$^b$}

\date{}

\maketitle

\vspace{-8mm}

\begin{center}
$^a${\it Department of Mathematics, Link\"oping University, S--581 83 Link\"oping,
Sweden \\ $^b$ Laboratory for Mathematical Modelling of Wave Phenomena, \\ Institute
for Problems in Mechanical Engineering, Russian Academy of Sciences, \\ V.O.,
Bol'shoy pr. 61, St. Petersburg 199178, RF} \\

\vspace{2mm}

E-mail: vladimir.kozlov@mai.liu.se ; nikolay.g.kuznetsov@gmail.com 
\end{center}

\begin{abstract}
The two-dimensional free-boundary problem describing steady gravity waves with
vorticity on water of finite depth is considered. It is proved that no
small-amplitude waves are supported by a horizontal shear flow whose free surface is
still in a coordinate frame such that the flow is time-independent in it. The class
of vorticity distributions for which such flows exist includes all positive constant
distributions, as well as linear and quadric ones with arbitrary positive
coefficients.

\vspace{2mm}

\end{abstract}

\section{Introduction}

\setcounter{equation}{0}

We consider the two-dimensional nonlinear problem of steady waves in a horizontal
open channel that has uniform rectangular cross-section and is occupied by an
inviscid, incompressible, heavy fluid, say, water. The water motion is assumed to be
rotational which, according to observations, is the type of motion commonly
occurring in nature (see, for example, \cite{SCJ,Th} and references cited therein).
A brief characterization of results obtained for this problem and a similar one
dealing with waves on water of infinite depth is given in \cite{KK2}. Further
details can be found in the survey article \cite{WS}.

In the present paper, our aim is to prove that no small-amplitude waves are
supported by a horizontal shear flow whose free surface is still in a coordinate
frame in which the flow is time-independent. In \cite{KK3}, all steady flows with
horizontal free surfaces are investigated in detail provided their stream functions
depend on the vertical coordinate only. Furthermore, the existence of Stokes waves
bifurcating from shear flows with non-still free surfaces is proved in \cite{KK2}
under rather natural assumptions, for which purpose a dispersion equation is
introduced and investigated. Thus, the results obtained here complement those in
\cite{KK2}. It is also worth mentioning that the case considered here and dealing
with the absence of waves essentially distinguishes from that when waves do not
arise on the free surface of the {\it critical} irrotational flow (see \cite{KK0},
Theorem~1 (i); the latter result complements the proof of the Benjamin--Lighthill
conjecture for the near-critical case obtained in \cite{KK,KK1}). Further details
concerning the hydrodynamic interpretation of the present result are given in \S 4.

As in the papers \cite{KK3,KK2}, no assumption is made about the absence of
counter-currents in a shear flow. Moreover, we impose no restriction on the type of
waves; they may be solitary, periodic with an arbitrary number of crests per period,
whatsoever. However, the slope of the free surface profile is supposed to be bounded
by a constant given {\it a priori}. Also, certain conditions which will be described
later are imposed on the vorticity distribution.

\subsection{Statement of the problem}

Let an open channel of uniform rectangular cross-section be bounded below by a
horizontal rigid bottom and let water occupying the channel be bounded above by a
free surface not touching the bottom. The surface tension is neglected and the
pressure is constant on the free surface. The water motion is supposed to be
two-dimensional and rotational which combined with the water incompressibility
allows us to seek the velocity field in the form $(\psi_y, -\psi_x)$, where $\psi
(x,y)$ is referred to as the {\it stream function} (see, for example, the book
\cite{LSh}). It is also supposed that the vorticity distribution $\omega$ (it is a
function of $\psi$ as is explained in \S 1 of the cited book) is a prescribed
Lipschitz function on $\RR$ subject to some conditions [see \eqref{eq:s_0} and
\eqref{eq:mu} below].

\vspace{2mm}

We use non-dimensional variables chosen so that the constant volume rate of flow per
unit span and the constant acceleration due to gravity are scaled to unity in our
equations. For this purpose lengths and velocities are scaled to $(Q^2/g)^{1/3}$ and
$(Qg)^{1/3}$, respectively; here $Q$ and $g$ are the dimensional quantities for the
rate of flow and the gravity acceleration, respectively. We recall that
$(Q^2/g)^{1/3}$ is the depth of the critical uniform stream in the irrotational
case (see, for example, \cite{Ben}).

In appropriate Cartesian coordinates $(x,y)$, the bottom coincides with the $x$-axis
and gravity acts in the negative $y$-direction. We choose the frame of reference so
that the velocity field is time-independent as well as the unknown free-surface
profile. The latter is assumed to be the graph of $y = \eta (x)$, $x\in \RR$, where
$\eta$ is a positive $C^{1}$-function. Therefore, the longitudinal section of the
water domain is $D = \{ x \in \RR,\ 0 < y < \eta (x) \}$, and $\psi$ is assumed to
belong to $C^2 (D) \cap C^{1} (\bar D)$.

\vspace{2mm}

Since the surface tension is neglected, $\psi$ and $\eta$ must satisfy the following
free-boundary problem:
\begin{eqnarray}
&& \psi_{xx} + \psi_{yy} + \omega (\psi) = 0, \quad (x,y)\in D; \label{eq:lapp} \\ &&
\psi (x,0) = 0, \quad x \in \RR; \label{eq:bcp} \\ && \psi (x,\eta (x)) = 1, \quad x
\in \RR; \label{eq:kcp} \\ && |\nabla \psi (x,\eta (x))|^2 + 2 \eta (x) = 3 r, \quad x
\in \RR . \label{eq:bep}
\end{eqnarray}
Here $r$ is a constant considered as the problem's parameter and referred to as the
total head (see, for example, \cite{KN}). This statement has long been known and
its derivation from the governing equations and the assumptions about the boundary
behaviour of water particles can be found, for example, in \cite{CS}.

Notice that the boundary condition \eqref{eq:kcp} yields that relation
\eqref{eq:bep} (Bernoulli's equation) can be written as follows:
\begin{equation} 
\left[ \partial_n \psi (x,\eta (x)) \right]^2  + 2 \eta (x) = 3 r, \quad x \in \RR
\, . \label{eq:ben}
\end{equation}
Here and below $\partial_n$ denotes the normal derivative on $\partial D$, and the
normal $n = (n_x, n_y)$ has unit length and points out of $D$.

\subsection{Assumptions and the result}

We begin with the conditions that are imposed on the vorticity distribution $\omega$
in our main theorem. Let $r_c$ denote the critical value of $r$ for $\omega$ (see
the definition in \cite{KK3}, p.~386; it is analogous to the total head of the
critical stream in the irrotational case). First, we require that
\begin{eqnarray} 
&& \!\!\!\!\!\!\!\!\!\!\!\!\!\! \mbox{for some $r > r_c$ problem
\eqref{eq:lapp}--\eqref{eq:bep} has a solution} \ (U (y), h) \ \mbox{with} \ h =
{\rm const} \nonumber \\ && \!\!\!\!\!\!\!\!\!\!\!\!\!\! \mbox{(such solutions are
referred to as {\it stream solutions}) for which} \ U_y (h) = 0 .
\label{eq:s_0}
\end{eqnarray}

Notice that if some pair $(\psi, \eta)$ satisfies problem
\eqref{eq:lapp}--\eqref{eq:bep} for the same $r$ as $(U , h)$, then equation
\eqref{eq:ben} for $(\psi, \eta)$ takes the form
\[ \left[ \partial_n \psi (x,\eta (x)) \right]^2  = 2 \, [ h - \eta (x) ] ,
\quad x \in \RR \, ,
\]
because $3 \, r = 2 \, h$ in this case. Hence $h - \eta (x) \geq 0$, which means
that if there exist a wavy flow perturbing the shear one of the depth $h$, then the
free surface of waves lies under the level $y = h$.

The second restriction that we impose on $\omega$ is as follows:
\begin{equation} 
\mu = \esssup_{\tau \in (-\infty, \infty)} \, \omega' (\tau) < \frac{\pi^2}{h^2} \, .
\label{eq:mu}
\end{equation}
This bound for $\mu$ is equal to the fundamental Dirichlet eigenvalue for the
operator $- \D^2 / \D^2 y$ on the interval $(0, h)$.

\vspace{2mm}

Now we are in a position to formulate our main result.

\vspace{2mm}

\noindent {\bf Theorem 1.} {\it Let the vorticity distribution $\omega$ satisfy
\eqref{eq:s_0} and \eqref{eq:mu}. Then for any $B > 0$ there exists $\varepsilon
(\mu, h, B) > 0$ such that every solution $(\psi, \eta)$ of problem
\eqref{eq:lapp}--\eqref{eq:bep} corresponding to the same $r$ as $(U , h)$ coincides
with the latter one provided}
\begin{equation} 
| \eta_x (x) | \leq B \quad and \quad h - \eta (x) < \varepsilon \quad for \ all \
x \in \RR \, . \label{eq:cond}
\end{equation}

The first and second inequalities \eqref{eq:cond} mean that the wave profile $\eta$
has bounded slope and sufficiently small amplitude, respectively.

\section{Auxiliary assertions}

Our proof of Theorem 1 is based on two lemmas. In the first of them, we estimate the
normal derivative of a solution satisfying an auxiliary boundary value problem in
the domain $D$. In the second lemma, some particular perturbation of the stream
function is estimated through the perturbation of the free surface profile. This
requires to reformulate the problem in terms of perturbations prior to formulating
lemmas.

\subsection{Reformulations of the problem}

First, we consider problem \eqref{eq:lapp}--\eqref{eq:bep} as a perturbation of that
for $(U, h)$ and write the problem for
\begin{equation}
\phi (x,y) = \psi (x,y) - U (y) \quad \mbox{and} \quad \zeta (x) = h - \eta (x) .
\label{eq:phi_zeta}
\end{equation}
Thus we obtain the following problem:
\begin{eqnarray}
&& \nabla^2 \phi + \omega (U + \phi) - \omega (U) = 0, \quad (x,y) \in D , \quad
\nabla = (\partial_x, \partial_y) ; \label{eq:lapp'} \\ && \phi (x,0) = 0, \quad x
\in \RR ; \label{eq:bcp'} \\ && \phi (x, h - \zeta (x)) = 1 - U (h - \zeta (x)),
\quad x \in \RR; \label{eq:kcp'} \\ && \left[ \partial_n \phi + \frac{U_y (y)}{(1 +
\zeta_x^2)^{1/2}} \right]^2_{y = h - \zeta (x)} = 2 \zeta (x), \quad x \in \RR .
\label{eq:bep'}
\end{eqnarray}
The last condition is a consequence of \eqref{eq:ben} and yields that $\zeta$ is a
non-negative function. 

In order to simplify condition \eqref{eq:kcp'}, we put
\[ w (x,y) = \phi (x,y) - u (x,y) , \quad \mbox{where} \ u (x,y) = [ 1 - 
U (h - \zeta (x)) ] \, \frac{y}{h - \zeta (x)} \, .
\]
The resulting problem for $w$ and $\zeta$ is as follows:
\begin{eqnarray}
&& \nabla^2 w + \omega (U + u + w) = \omega (U) - \nabla^2 u , \quad (x,y) \in D ;
\label{eq:lapp"} \\ && w (x,0) = 0, \quad x \in \RR ; \label{eq:bcp"} \\ 
&& w (x, h - \zeta (x)) = 0 , \quad x \in \RR; \label{eq:kcp"} \\ && \left[
\frac{\partial_n w}{(1 + \zeta_x^2)^{1/2}} + \frac{1 - U (y)}{y} + U_y (y)
\right]^2_{y = h - \zeta (x)} = \frac{2 \zeta (x)}{1 + \zeta_x^2} \, , \quad x \in
\RR . \label{eq:bep"}
\end{eqnarray}

In conclusion, we list a couple of properties that will be used below. If $\zeta$
is small enough, then the inequalities
\begin{equation}
|u (x,y)| \leq C \, [\zeta (x)]^2 , \quad |u_y (x,y)| \leq C \, [\zeta (x)]^2 ,
\quad |u_x (x,y)| \leq C \, |\zeta_x (x)| \, \zeta (x) \label{eq:u}
\end{equation}
immediately follow from the definition of $u$. In the first and second of them, the
constant $C$ depends on the stream solution $(U, h)$, whereas the constant is
absolute in the last inequality. Hence the conditions imposed on $\omega$ yield that
$|w|$ is bounded on $\bar D$.

\subsection{Two lemmas}

For an arbitrary $t \in \RR$ we define the following truncated domain:
\[ D_t = \{ (x,y) : x \in (t-1, t+2) , \, y \in (0 , \eta (x) \} . \]


\noindent {\bf Lemma 1.} {\it Let $y = \eta (x)$ be a fixed curve such that the first condition \eqref{eq:cond} is
fulfilled. Let also $\eta (x) \geq h_-$ for all $x$, where $h_-$ is some positive
constant. If $v$ is a solution of the problem
\[ \nabla^2 v = f , \ \ (x,y) \in D ; \quad v (x,0) = 0 , \ \ x \in \RR ; \quad 
v (x, \eta (x)) = H , \ \ x \in \RR 
\] 
with $f \in L^{2}_{loc} (D)$ and $H \in W^{1,2}_{loc} (\RR)$, then for every $t \in
\RR$ the following estimate holds:
\begin{equation}
\left\| \partial_n v |_{y=\eta (x)} \right\|_{L^2 (t, t+1)} \leq C \left[ \| f
\|_{L^2 (D_t)} + \| H \|_{W^{1,2} (t-1, t+2)} + \| v \|_{W^{1,2} (D_t)} \right] ,
\label{eq:lem_1}
\end{equation}
where the constant $C$ does not depend on $f$, $H$ and $t$.}

\vspace{2mm}

\noindent {\it Proof.} By $\chi$ we denote a smooth cut-off function such that
$\chi (x) = 1$ for $x \in (t, t+1)$, $\chi (x) = 0$ for $x \in (-\infty, t-1/2) \cup
(t+3/2, +\infty)$ and $0 \leq \chi (x) \leq 1$ for all $x$. Let us multiply the
equality
\[ \nabla^2 (\chi v) = \chi f + v \nabla^2 \chi + 2 \, \nabla v \cdot \nabla \chi \]
by $(\chi v)_y$ and integrate over $D$, thus obtaining
\begin{eqnarray}
-\frac{1}{2} \int_D \left( | \nabla (\chi v) |^2 \right)_y \D x \D y +
\int_{\partial D} (\chi v)_y \, \partial_n (\chi v) \, \D s \nonumber \\ = \int_D
\left( \chi f + v \nabla^2 \chi + 2 \, \nabla v \cdot \nabla \chi \right) (\chi v)_y
\, \D x \D y \, . \label{eq:lem_1.1}
\end{eqnarray}
The expression in the left-hand side arises after applying the first Green's
formula; $\D s$ stands for element of the arc length. Introducing $\partial_t$ so
that $\nabla = (\partial_t, \partial_n)$ on $y = \eta (x)$, we transform the
left-hand side as follows:
\begin{eqnarray*}
&& -\frac{1}{2} \int_{-\infty}^{+\infty} \left[ | \nabla (\chi v) |^2
\right]_{y=0}^{y = \eta (x)} \D x + \int_{-\infty}^{+\infty} \left[ (n_y \partial_n
- n_x \partial_t) (\chi v) \, \partial_n (\chi v) \right]_{y = \eta (x)}
\sqrt{1+\eta_x^2} \, \D x \\ && - \int_{-\infty}^{+\infty} \left[ (\chi v)_y^2
\right]_{y = 0} \D x = \int_{-\infty}^{+\infty} \left( n_y \sqrt{1+\eta_x^2} -
\frac{1}{2} \right) \left[ \partial_n (\chi v) \right]_{y = \eta (x)}^2 \D x \\ && -
\int_{-\infty}^{+\infty} \left[ n_x \sqrt{1+\eta_x^2} \, \partial_t (\chi v) \,
\partial_n (\chi v) + \frac{1}{2} \left| \partial_t (\chi v) \right|^2 \right]_{y =
\eta (x)} \D x - \int_{-\infty}^{+\infty} \left[ (\chi v)_y^2 \right]_{y = 0} \D x
.
\end{eqnarray*}
We substitute the last expression into \eqref{eq:lem_1.1} and take into account that
$n_x \sqrt{1+\eta_x^2} = -\eta_x$, whereas the first factor in the first integrand is
equal to $1/2$. Then we arrive, after re\-arranging terms and multiplying by two, at
the following equality:
\begin{eqnarray*}
\int_{-\infty}^{+\infty} \left[ \partial_n (\chi v) \right]_{y = \eta (x)}^2 \D x =
\int_{-\infty}^{+\infty} \left[ \left| \partial_t (\chi v) \right|^2 - 2 \, \eta_x
\, \partial_t (\chi v) \, \partial_n (\chi v) \right]_{y = \eta (x)} \D x \\ +
\int_{-\infty}^{+\infty} \left[ (\chi v)_y^2 \right]_{y = 0} \D x + \int_D \left(
\chi f + v \nabla^2 \chi + 2 \, \nabla v \cdot \nabla \chi \right) (\chi v)_y \, \D
x \D y \, .
\end{eqnarray*}
Since the left-hand side in \eqref{eq:lem_1} is less than that in the last equality,
it is sufficient to estimate with proper constants each term in the right-hand side
in order to complete the proof of the required inequality \eqref{eq:lem_1}.

First, we have that
\[ \left| \int \limits_{-\infty}^{+\infty} \left[ \eta_x \, \partial_t (\chi v) 
\, \partial_n (\chi v) \right]_{y = \eta (x)} \D x \right| \leq \frac{1}{4} \int
\limits_{-\infty}^{+\infty} \left[ \partial_n (\chi v) \right]_{y = \eta (x)}^2 \D x
+ 4 \, B^2 \int \limits_{-\infty}^{+\infty} \left[ \partial_t (\chi v) \right]_{y =
\eta (x)}^2 \D x \, ,
\]
because $y = \eta (x)$ satisfies the first condition \eqref{eq:cond}. Furthermore,
the assumption that $\eta (x) \geq h_-$ for all $x$, where the constant $h_- > 0$,
allows us to apply the general theory of elliptic boundary value problems (see, for
example, \cite{ADN}), from which it follows that
\[ \int_{-\infty}^{+\infty} \left[ (\chi v)_y^2 \right]_{y = 0} \D x \leq C \left[ 
\| f \|_{L^2 (D_t)} + \| H \|_{W^{1,2} (t-1, t+2)} + \| v \|_{W^{1,2} (D_t)} \right]
,
\]
where $C$ depends only on $h_-$. Finally, using the Schwarz and Cauchy inequalities,
one readily obtains that the absolute value of the integral over $D$ is estimated by
the right-hand side in the last inequality. \hfill $\bigcirc$

\vspace{2mm}

Applying lemma 1 to problem \eqref{eq:lapp'}--\eqref{eq:kcp'} (we are able to do
this because $\omega$ is locally Lipschitz and has bounded derivative), we obtain
the following.

\vspace{2mm}

\noindent {\bf Corollary 1.} {\it If $\phi$ is defined by the first formula \eqref{eq:phi_zeta}, then the estimate
\eqref{eq:lem_1} for $\phi$ takes the form:
\[ \left\| \partial_n \phi |_{y= h - \zeta (x)} \right\|_{L^2 (t, t+1)} \leq C \left[ 
\| \phi \|_{W^{1,2} (D_t)} + \| 1 - U (h - \zeta) \|_{W^{1,2} (t-1, t+2)} \right] .
\]
Moreover, the last term in the square brackets does not exceed $(\varepsilon + B) \|
\zeta \|_{L^2 (t, t+1)}$ provided conditions \eqref{eq:cond} are fulfilled.}

\vspace{2mm}

\noindent {\bf Lemma 2.} {\it Let the conditions imposed on $\omega$, $r$ and $(U,
h)$ in theorem~1 be fulfilled. If $\zeta$ is sufficiently small and $|\zeta_x| \leq
B$ for some $B > 0$, then there exist $\delta > 0$, depending on $(\pi/h)^2 - \mu$,
$h$ and $B$, and $C_\delta > 0$ such that the following inequality
\begin{equation} 
\int_D \E^{-\delta |t-x|} \left( w^2 + |\nabla w|^2 \right) \D x \D y \leq C_\delta
\int_{-\infty}^{+\infty} \E^{-\delta |t-x|} \zeta^2 \left( \zeta^2 + \zeta^2_x
\right) \D x \label{eq:lem_2}
\end{equation}
holds for every function $w$ satisfying relations \eqref{eq:lapp"}--\eqref{eq:kcp"}
and all $t \in \RR$.}

\vspace{2mm}

\noindent {\it Proof.} Let $\chi_N (x)$ denote a cut-off function equal to unity on
$(-N, N)$ and vanishing for $|x| > 2 \, N$. We write equation \eqref{eq:lapp'} in
the form
\[ \nabla^2 w + \omega (U + u + w) - \omega (U + u) = \omega (U) - \omega (U + u)
- \nabla^2 u ,
\]
multiply it by $- w (x) \, \chi_N (x - t) / \cosh \delta (x - t)$ with some $\delta
> 0$, and integrate over $D$. After applying the first Green's formula and
integrating by parts in the left-hand side, we arrive at the following equality:
\begin{eqnarray}
&& \int_D \left\{ \frac{\chi_N (x - t)}{\cosh \delta (x - t)} \left( |\nabla w|^2 -
w \int_0^w \omega' (U + u + \tau) \, \D \tau \right) - \frac{w^2}{2} \left[
\frac{\chi_N (x - t)}{\cosh \delta (x - t)} \right]_{xx} \right\} \D x \D y
\nonumber \\ && \ \ \ \ \ \ \ \ \ \ \ \ \ \ \ \ \ \ \ \ \ \ \ \ \ \ \ \ \ \ \ \ \  =
\int_D \frac{\chi_N (x - t)}{\cosh \delta (x - t)} \, w \left[ \nabla^2 u + \omega
(U + u) - \omega (U) \right] \D x \D y \, . \label{eq:pr_1}
\end{eqnarray}
Here the boundary conditions \eqref{eq:bcp"} and \eqref{eq:kcp"} are also taken into
account. 

Using assumption \eqref{eq:mu}, we get that the absolute value of the left-hand
side is greater than or equal to
\begin{eqnarray}
&& \!\!\!\!\!\!\!\!\!\!\!\! \int_D \bigg\{ \frac{\chi_N (x - t)}{\cosh \delta (x -
t)} \left[ |\nabla w|^2 - (\mu + 3 \, \delta^2) \, w^2 \right] \nonumber \\ && -
\frac{w^2}{2} \left| \frac{\chi_N'' (x - t)}{\cosh \delta (x - t)} + 2 \, \chi_N' (x
- t) \left[ 1 / \cosh \delta (x - t) \right]' \right| \bigg\} \, \D x \D y \, ,
\label{eq:pr_2}
\end{eqnarray}
because $\left| \left( 1 / \cosh \delta x \right)'' \right| \leq 3\, \delta^2 /
\cosh \delta x$. Furthermore, we have that
\[ \int_0^{h-\zeta} w_y^2 \, \D y \geq \delta^2 \int_0^{h-\zeta} w_y^2 \, \D y +
(1 - \delta^2) \, (\pi / h)^2 \int_0^{h-\zeta} w^2 \, \D y \, ,
\]
which gives that the integral in the first line of \eqref{eq:pr_2} is estimated from
below by the following expression:
\begin{equation} 
\int_D \frac{\chi_N (x - t)}{\cosh \delta (x - t)} \left\{ \left( w_x^2 + \delta^2
w_y^2 \right) + \left[ (1 - \delta^2) \, (\pi / h)^2 - \mu - 3 \, \delta^2 \right]
\, w^2 \right\} \D x \D y \, . \label{eq:pr_3}
\end{equation}
In view of assumption \eqref{eq:mu}, the number in the square brackets is positive
provided $\delta$ is chosen sufficiently small.

Now we turn to estimating from above the absolute value of the right-hand side in
\eqref{eq:pr_1}. First, the Cauchy inequality yields that
\begin{eqnarray}
\left| \int_D \frac{\chi_N (x - t)}{\cosh \delta (x - t)} \left[ \omega (U + u) -
\omega (U) \right] \D x \D y \right| \leq C_\omega \int_D \frac{\chi_N (x -
t)}{\cosh \delta (x - t)} \, |u \, w| \, \D x \D y \nonumber \\ \leq \delta^2 \int_D
\frac{\chi_N (x - t)}{\cosh \delta (x - t)} \, w^2 \, \D x \D y + \frac{C_\omega^2}{4
\delta^2} \int_D \frac{\chi_N (x - t)}{\cosh \delta (x - t)} \, u^2 \, \D x \D y ,
\label{eq:pr_4}
\end{eqnarray}
where $C_\omega$ is the Lipschitz constant of $\omega$. Second, we apply the first
Green's formula to the other term and get, in view of the boundary conditions
\eqref{eq:bcp"} and \eqref{eq:kcp"}, that its absolute value can be written as
follows:
\begin{equation} 
\left| \int_D \left\{ \frac{\chi_N (x - t)}{\cosh \delta (x - t)} \nabla u \cdot
\nabla w + w u_x \left[ \frac{\chi_N' (x - t)}{\cosh \delta (x - t)} + \chi_N (x -
t) \left[ \frac{1}{\cosh \delta (x - t)} \right]' \right] \right\} \D x \D y \right|
. \label{eq:pr_7} 
\end{equation}
Here the first and third terms do not exceed
\begin{equation} 
\frac{\delta^2}{2} \int_D \frac{\chi_N (x - t)}{\cosh \delta (x - t)} \, |\nabla
w|^2 \, \D x \D y + \frac{1}{2 \, \delta^2} \int_D \frac{\chi_N (x - t)}{\cosh
\delta (x - t)} \, |\nabla u|^2 \, \D x \D y \label{eq:pr_5}
\end{equation}
and
\begin{equation} 
\delta^2 \int_D \frac{\chi_N (x - t)}{\cosh \delta (x - t)} \, w^2 \, \D x \D y +
\frac{1}{4} \int_D \frac{\chi_N (x - t)}{\cosh \delta (x - t)} \, u_x^2 \, \D x \D y
\, , \label{eq:pr_6}
\end{equation}
respectively, whereas we simply take the absolute value of the integrand in the
second term.

Using \eqref{eq:pr_2}--\eqref{eq:pr_6} in equality \eqref{eq:pr_1} and letting $N
\to \infty$, we arrive at the following inequality:
\begin{eqnarray*}
&& \int_D \left[ \left( 1 - \frac{\delta^2}{2} \right) w_x^2 + \frac{\delta^2}{2}
w_y^2 + \left\{ \left( \frac{\pi}{h} \right)^2 - \mu - \delta^2 \left[ 5 + \left(
\frac{\pi}{h} \right)^2 \right] \right\} w^2 \right] \frac{\D x \D y}{\cosh \delta
(x - t)} \\ && \ \ \ \ \ \ \ \ \ \ \ \ \ \ \ \ \ \ \ \ \ \ \ \ \ \ \ \ \ \ \ \leq
\int_D \left[ \left( \frac{1}{4} + \frac{1}{2 \, \delta^2} \right) |\nabla u|^2 +
\frac{C_\omega^2}{4 \, \delta^2} \, u^2 \right] \frac{\D x \D y}{\cosh \delta (x - t)} ,
\end{eqnarray*}
because $\chi_N$ goes to unity, whereas $\chi_N'$ and $\chi_N''$ go to zero. Now
\eqref{eq:lem_2} follows from the definition of $u$ and assumption \eqref{eq:mu}.
\hfill $\bigcirc$

\vspace{2mm}

A consequence of lemma 2 is the following.

\vspace{2mm}

\noindent {\bf Corollary 2.} {\it Let the assumptions of lemma 2 be fulfilled, and
let $\zeta (x) < h$ for all $x \in \RR$. Then}
\[ \| w \|_{W^{1,2} (D_t)} \leq C (\delta, h, B) \sup_{\tau \in \RR} 
\| \zeta \|_{L^2 (\tau, \tau+1)} \quad for \ all \ t \in \RR .
\]

\vspace{2mm}

\noindent {\it Proof.} It is clear that the left-hand side of \eqref{eq:lem_2}
is greater than or equal to
\[ \int_{t-1}^{t+2} \E^{-\delta |t-x|} \, \D x \int_0^{h-\zeta} \left( w^2 + |\nabla
w|^2 \right) \D y \geq \E^{-2 \delta} \, \| w \|_{W^{1,2} (D_t)} , 
\]
because $\E^{-2 \delta} = \min_{x \in [t-1, t+2]} \E^{-\delta |t-x|}$. Since
\[ \int_{-\infty}^\infty f (x) \, \D x = \int_{-\infty}^\infty \D \tau \int_\tau^{\tau
+ 1} f (x) \, \D x \quad \mbox{for any} \ f ,
\]
we write the right-hand side of \eqref{eq:lem_2} as follows:
\[ C_\delta \int_{-\infty}^{+\infty} \D \tau \int_\tau^{\tau
+ 1} \E^{-\delta |t-x|} \, \zeta^2 \left( \zeta^2 + \zeta^2_x \right) \D x \, .
\]
This, in view of the assumptions made about $\zeta$ and $\zeta_x$, is less than or
equal to
\[ C_\delta \, \E^{\delta} (h^2 + B^2) \int_{-\infty}^{+\infty} \E^{-\delta |t-\tau|}
\, \| \zeta \|_{L^2 (\tau, \tau + 1)} \D \tau \, ,
\]
because $\E^{-\delta |t-x|} \leq \E^{\delta} \E^{-\delta |t-\tau|}$ provided $\tau
\leq x \leq \tau + 1$. Taking the supremum of the norm, we arrive at the required
inequality, because the integral of $\E^{-\delta |t-\tau|}$ is equal to $2 /
\delta$. \hfill $\bigcirc$

\section{Proof of theorem 1}

The assumptions made about $\eta$ and $\eta_x$ allows us
to apply inequalities \eqref{eq:u} for estimating $u$ and corollary~2 for estimating
$w$. Since $\phi = u + w$, we get
\[ \| \phi \|_{W^{1,2} (D_t)} \leq C (B^2 + h^2)^{1/2} \| \zeta \|_{L^2 (t-1, t+2)} 
+ C (\delta, h, B) \sup_{\tau \in \RR} \| \zeta \|_{L^2 (\tau, \tau+1)} ,
\] 
and so the right-hand side does not exceed $C_1 (\delta, h, B) \sup_{\tau \in \RR}
\| \zeta \|_{L^2 (\tau, \tau+1)}$. Combining this fact and corollary~1, we obtain
that
\begin{equation} 
\left\| \partial_n \phi |_{y= h - \zeta (x)} \right\|_{L^2 (t, t+1)} \leq C_2
(\delta, h, B) \sup_{\tau \in \RR} \| \zeta \|_{L^2 (\tau, \tau+1)} \leq
\varepsilon^{1/2} C_2 (\delta, h, B) \sup_{\tau \in \RR} \| \zeta \|_{L^1 (\tau,
\tau+1)}^{1/2} , \label{eq:th}
\end{equation}
where the last inequality is a consequence of the second assumption
\eqref{eq:cond}.

Bernoulli's equation written as follows [cf. \eqref{eq:bep'}]
\[ [\zeta (x)]^{1/2} = \frac{1}{\sqrt 2} \left| \partial_n \phi + \frac{U_y (y)}
{(1 + \zeta_x^2)^{1/2}} \right|_{y = h - \zeta (x)} , \quad x \in \RR ,
\]
immediately yields that
\[ \sup_{\tau \in \RR} \| \zeta \|_{L^1 (\tau, \tau+1)}^{1/2} \leq \frac{1}{\sqrt 2}
\sup_{\tau \in \RR} \left[ \left\| \partial_n \phi |_{y= h - \zeta (x)}
\right\|_{L^2 (\tau, \tau+1)} + C \, \| \zeta \|_{L^2 (\tau, \tau+1)} \right] .
\]
Using inequalities \eqref{eq:th} for estimating both terms in the square brackets,
we arrive at
\[ \sup_{\tau \in \RR} \| \zeta \|_{L^1 (\tau, \tau+1)}^{1/2} \leq \varepsilon^{1/2} 
C \sup_{\tau \in \RR} \| \zeta \|_{L^1 (\tau, \tau+1)}^{1/2} \, ,
\]
which is impossible for sufficiently small $\varepsilon$. The obtained contradiction
proves theorem~1.

\section{Discussion}

In the framework of the classical approach to steady water waves with vorticity, it
is proved under assumptions \eqref{eq:s_0} and \eqref{eq:mu} that no waves of small
amplitude are supported by a horizontal shear flow with still free surface. Here we
discuss the first of these assumptions in greater detail and consider examples when
both of them are fulfilled.

\vspace{2mm}

The first assumption (there exists a stream solution with still free surface)
yields that
\[ h_0 = \int_0^1 \frac{\D \tau}{\sqrt{s_0^2 - 2 \, \Omega (\tau)}} < \infty , \
\mbox{where} \ \Omega (\tau) = \int_0^\tau \omega (t) \, {\D} t \ \mbox{and} \ s_0
= \sqrt{2 \, \max_{\tau \in [0,1]} \Omega (\tau)} \, . 
\]
Let the maximum is attained at $\tau_0 \in [0,1]$, then $h_0 < \infty$ if and only
if $\omega (\tau_0) \neq 0$, and so $\tau_0$ is either 0 or 1. These are the
conditions of either case (ii) or case (iii), according to the classification of
vorticity distributions (see \cite{KK3}, \S 4.2).

For $s_0 > 0$ any stream solution $(U , h)$ that satisfies assumption \eqref{eq:s_0}
is as follows:
\begin{equation} 
\mbox{either} \ \ \left( U (y ; s_0) , h_k^{(+)} \right) \ \ \mbox{or} \ \ \left( U
(y ; -s_0) , h_k^{(-)} \right) . \label{eq:+/-}
\end{equation}
Here $U (y ; s)$ denotes (as in the cited paper) a unique solution of the Cauchy 
problem:
\[
U_{yy} + \omega (U) = 0 , \quad U (0) = 0 , \quad U_y (0) = s , 
\]
whereas $h_k^{(+)} = h_0 + 2 \, k \, [ h_0 - y_- (s_0) ]$ and $h_k^{(-)} = h_k^{(+)}
- 2 \, y_- (s_0)$, $k = 0,1,\dots$; $y_- (s_0)$ is such that $(y_- (s_0) , h_0)$ is
the maximal interval containing $y=0$ inside, on which $U (y ; s_0)$ increases
strictly monotonically. Thus, if $y_- (s_0) > -\infty$, then $U (y ; s_0)$ is
periodic and the above formulae are valid for all non-negative integers $k$.
Otherwise, only the first formula \eqref{eq:+/-} with $k=0$ gives a stream solution
satisfying assumption \eqref{eq:s_0}.

For $s_0 = 0$ we have $y_- (s_0) = 0$, and so all stream solutions satisfying
assumption \eqref{eq:s_0} are given by the first formula \eqref{eq:+/-} provided $U
(y ; s_0)$ is periodic.

\vspace{2mm}

Now we turn to examples of vorticity distributions $\omega$ for which both
assumptions \eqref{eq:s_0} and \eqref{eq:mu} are fulfilled.

First, we take the vorticity equal to an arbitrary positive constant, say, $b > 0$
(see details in \cite{KK3}, \S 6.1), and obtain the simplest example of the unique
stream solution satisfying \eqref{eq:s_0} and \eqref{eq:mu} simultaneously. Indeed,
in this case $s_0 = \sqrt{2 \, b} > 0$, $h_0 = \sqrt{2 / b} = h$ and the stream
function is $U = \sqrt{2 \, b} \, y - b \, y^2 / 2$. Therefore, the corresponding
shear flow has the velocity profile in the form of a straight segment which goes
from $\sqrt{2 \, b}$ on the bottom to zero on the free surface. In his study
\cite{Wah} of bifurcation of waves from shear flows with constant vorticity,
Wahl\'en also excluded the above stream solution from his considerations.

On the contrary, if the vorticity is equal to a negative constant, say, $b < 0$,
then $s_0 = 0$, and the corresponding stream solution $(U, h) = (b \, y^2 / 2 ,
\sqrt{2 / b})$ gives a positive value of the flow velocity on the free surface. The
existence of Stokes waves bifurcating from this shear flow is proved in \cite{Wah},
but the general results obtained in \cite{KK2} are not applicable in this case.
Presumably, the reason for this lies in the degeneration of the streamline pattern
for $s_0 = 0$, which becomes clear from figures~1 and 2 in \cite{Wah}. Indeed, the
velocity of flow is negative (vanishes) on the bottom for the flow shown in figure~1
(figure~2, respectively). In the middle of the flow corresponding to the negative
bottom velocity (see figure~1), there is a critical layer formed by closed cat's-eye
vortices. However, for $s_0 = 0$ domains with closed streamlines are attached to the
bottom and separated from each other.

\vspace{2mm}

In the case of positive linear vorticity, that is, $\omega (\tau) = b \, \tau$, $b >
0$, we have that $s_0 = \sqrt{b}$ and $h_0 = \pi / (2 \sqrt{b})$ (see details in
\cite{KK3}, \S 6.3). There are infinitely many stream solutions corresponding to
$s_0$, and their second components are equal to 
\[ \frac{\pi \, (2 \, k - 1)}{2 \sqrt{b}} \quad (k=1,2,\dots) .
\]
Condition \eqref{eq:mu} is fulfilled only for $k = 1$, in which case theorem~1 is
valid, but it gives no answer for $k \geq 2$. However, all shear flows with still
free surfaces are excluded from consideration in the detailed study \cite{EEW} of
waves with positive linear vorticity. The reason for this is as follows: `without
this assumption the linearized operator [...] appearing in the bifurcation problem'
can be shown not to be Fredholm.

\vspace{2mm}

Theorem~1 is also applicable to a shear flow with $\omega (\tau) = b \, \tau^2$ on
$[-R, R]$ and constant $\omega (\tau)$ for $|\tau|$ outside $(-R, R)$ (the constant
is taken so that $\omega$ is continuous); here $R > 1$ and $b$ is a positive
constant. For this vorticity we have $s_0 = \sqrt{2 \, b / 3}$, whereas formula
\eqref{eq:s_0} gives that
\[ h_0 = \sqrt{\frac{3}{2 \, b}} \int_0^1 \frac{\D \tau}{\sqrt{1 - \tau^3}} \, . \]
The equation for the first component of the corresponding stream solution is as
follows: 
\[ 3 \, U_y^2 + 2 \, b \, U^3 = 2 \, b .
\]
Using elliptic functions, one can obtain its general solution (see \cite{Kam},
pt.~3, ch.~6, \S 6.5), but this is superfluous in the present context. Of course,
the smallest (if there are more than one) second component of stream solutions with
still free surfaces is equal to $h_0$ for which, according to formula 17.4.59 in
\cite{AS}, we have the following expression:
\[ \sqrt{\frac{3}{2 \, b}} \frac{F (\varphi_0 \backslash \alpha_0)}{\sqrt[4]{3}} 
\, , \quad \mbox{where} \ \varphi_0 = \arccos \frac{\sqrt 3 - 1}{\sqrt 3 + 1} \, , \
\alpha_0 = 75^\circ ,
\] 
and $F (\varphi \backslash \alpha)$ denotes elliptic integral of the first kind.
Then condition \eqref{eq:mu} is fulfilled if $\sqrt 3 \, [ F (\varphi_0 \backslash
\alpha_0) ]^2 < \pi^2$, and this inequality is true because after simple
computations one gets from table 17.5 in \cite{AS} that $F (\varphi_0 \backslash
\alpha_0) < 1.9$.

\vspace{2mm}

Any of the described above shear flows might be called a {\it critical flow of the
second kind}. Indeed, Stokes waves bifurcate from all shear flows whose depths are
close to $h$ for positive constant and positive linear vorticity (see \cite{KK2}, \S
5). On the other hand, the bifurcation pattern is different near a flow that is
referred to as {\it critical} on p.~386 of the cited paper. We recall that this flow
described by $(U (y; s_c) , h (s_c))$ exists for {\it all} vorticity distributions.
On the $s$-axis, the value $s_c$ separates two intervals with different properties.
On the left of $s_c$, there lies a finite interval and for $s$ belonging to it
small-amplitude Stokes waves bifurcate from the corresponding horizontal shear flows
(see Main Theorem in \cite{KK2}). On the right of $s_c$, a sufficiently small
interval exists such that solitary waves are present for those $s$ as is proved in
\cite{H1}. This near-critical behaviour distinguishes from that outlined above, but
is completely analogous to that taking place in the irrotational case when the
critical uniform flow separates sub- and supercritical flows from which Stokes and
solitary waves, respectively, bifurcate (see, for example, \cite{KK,KK1}). Besides,
only a uniform flow exists for the critical value of the problem's parameter in the
irrotational case (see \cite{KK0}, Theorem~1, for the proof). On the contrary, a
similar fact for problem \eqref{eq:lapp}--\eqref{eq:bep} is still an open question.

\vspace{6mm}

\noindent {\bf Acknowledgements.} V.~K. was supported by the Swedish Research
Council (VR). N.~K. acknowledges the financial support from the Link\"oping
University.

\end{document}